\documentclass[aps,prb,twocolumn,showpacs]{revtex4}
\usepackage{graphicx,amssymb,amsmath}
\pdfoutput=1

\begin{document}

\title{Spin-selective localization due to intrinsic spin-orbit coupling}

\author{Yaroslav Tserkovnyak}
\author{Shimul Akhanjee}
\affiliation{Department of Physics and Astronomy, University of California, Los Angeles, California 90095, USA}

\date{\today}

\begin{abstract}
We study spin-dependent diffusive transport in the presence of a tunable spin-orbit (SO) interaction in a two-dimensional  electron system. The spin precession of an electron in the SO coupling field is expressed in terms of a covariant curvature, affecting the quantum interference between different electronic trajectories. Controlling this curvature field by modulating the SO coupling strength and its gradients by,  e.g., electric or elastic means, opens intriguing possibilities for exploring spin-selective localization physics. In particular, applying a weak magnetic field allows the control of the electron localization independently for two spin directions, with the spin-quantization axis that could be ``engineered" by appropriate SO interaction gradients.
\end{abstract}



\maketitle

\section{Introduction}

Recent years have seen a growing interest in spin-orbit (SO) interactions in metals, semiconductors, and topological insulators, with significant advances in the physics of spin Hall effects and other phenomena, where SO coupling profoundly modifies electronic transport.\cite{zuticRMP04,engelCHA07,tserkovPRB07sh,tserkovPRB07ac} The interest in these problems is also fueled by the desire to develop spintronic circuits that do not rely on magnetic elements.

Spin-transport properties are in general associated with a gauge freedom of position-dependent spin rotation. This SU(2) gauge symmetry and the associated curvature field were recently discussed in the contexts of the spin Hall physics, the Aharonov-Casher effect, and other spin-transport phenomena.\cite{jinJPA06} Additionally, there have been parallel studies on laser-induced non-Abelian gauge fields in cold-atom optical lattices,\cite{osterlohPRL05} which discussed the Hofstadter ``moth" spectrum, the Anderson localization transition, and the possibility of realizing non-Abelian interferometry. Understanding the structure of the gauge-covariant curvature field underlying the SO interaction may thus carry fundamental and broad implications.

In this article, we explore the physical consequences that arise from topological structure of this SU(2) gauge field. As a practical example, we will consider quantum transport corrections in a disordered two-dimensional (2D) conductor with intrinsic SO coupling. In particular, by focusing on the Rashba interaction we demonstrate the possibility of a tunable spin-selective localization of the charge carriers, by spatially modulating the strength of the SO interaction, which can be achieved via the careful control of electric gates or by elastic strain.

Our formalism is based on the geometric nature of the precession of electrons in the presence of SO fields, which will facilitate the discussion of the weak-localization effects: The underlying quantum interference between closed time-reversed trajectories can be analytically described by a path integral over a non-Abelian gauge field or Wilson-loop integral. We find that the effective size of these pairs of trajectories provides a natural limit for separating Abelian and non-Abelian contributions. A compact, Abelian treatment using the Stokes' theorem can be applied to smaller loops with a moderate net spin precession, while a large spin precession is responsible for the destruction of the phase coherence of longer time-reversed loops, where the non-Abelian contributions play a dominant role. Our discussion is relevant to recent experimental studies of weak (anti)localization in 2D heterostructures with strong intrinsic SO coupling,\cite{kogaPRL02} and is complementary to the established theoretical framework for disordered systems,\cite{knapPRB96} by explicitly revealing the topological structure and exploring the role of SO gradients.

In the following, we will start by formally casting the SO interaction as a Yang-Mills gauge field and  the interference effects in terms of the gauge-covariant curvature field, using the generalized Stokes' theorem. This SU(2) curvature can be diagonalized in spin space, with its eigenvalues defining a spin-dependent fictitious magnetic field. The corresponding spin-quantization axis has in-plane and out-of-plane components, that depend on both the strength and gradients of the SO coupling. Therefore, the SO interaction combined with an external magnetic field generates a total effective field that can be tuned independently for spins up and down along a desirable spin-quantization axis. These ideas can be applied towards a semiclassical description of SO effects on the weak-localization physics in 2D, providing a simpler and physically more intuitive picture than offered by the conventional diagrammatic Cooperon perspective. Our approach is especially advantageous when dealing with system having inhomogeneous SO interactions.

\section{model}
\subsection{Effective SU(2) gauge field}

Consider the Hamiltonian for 2D electrons in the presence of both electric and magnetic fields. An SO interaction arises from the coupling of an electron's spin with its orbital motion, which stems from relativistic corrections as electron accelerates in the presence of crystalline or applied electric field. Although there can be many different forms of the SO interaction, the primary difference originates from the type of electric field source. For example, when the source electric field is from an in-plane impurity potential, the type of SO interaction is called ``extrinsic." In contrast, an applied electric field, elastic strain, crystalline inversion asymmetry, or 2D confinement asymmetry perpendicular to the two-dimensional electron gas (2DEG) results in an ``intrinsic" SO interaction incapsulated in the following Hamiltonian:
\begin{equation}
\hat{H}=\frac{1}{2m}\left(-i\hbar\boldsymbol{\nabla}+\frac{e}{c}\hat{\mathbf{A}}(\mathbf{r})\right)^2-eV(\mathbf{r})\,,
\label{H}
\end{equation}
where $\hat{\mathbf{A}}(\mathbf{r})$ is an atomically smooth function. $m$ is the effective mass, $-e$ is the electron charge, and $V(\mathbf{r})$ is a scalar in-plane, quenched impurity potential. Hamiltonian (\ref{H}) follows from the effective-mass approximation, and we retained the general linear in momentum spin-orbit interaction, in terms of the $2\times2$ ``vector potential" $\hat{\mathbf{A}}$, or ``connection" with noncommuting matrix components. The associated  ``curvature" is described by the covariant tensor field:\cite{brodaCHA01}
\begin{equation}
\hat{\mathcal{F}}_{\mu\nu}=\partial_\mu\hat{A}_\nu-\partial_\nu\hat{A}_\mu+\frac{2\pi i}{\phi_0}[\hat{A}_\mu,\hat{A}_\nu]\,,
\label{F}
\end{equation}
where $\mu$ and $\nu$ label the two spatial components, and we have, for simplicity, neglected Zeeman interactions (which would produce also the time component of the connection field). $\phi_0=hc/e$ is the magnetic flux quantum. It is easy to verify that a position-dependent SU(2) wave-function transformation corresponding to Hamiltonian (\ref{H}), $\hat{\psi}=\hat{U}(\mathbf{r})\hat{\psi}^\prime$, transforms the connection as $\hat{A}^\prime_\mu=\hat{U}^\dagger\hat{A}_\mu\hat{U}-(i\phi_0/2\pi)\hat{U}^\dagger\partial_\mu\hat{U}$, and the associated curvature simply as $\hat{\mathcal{F}}^\prime_{\mu\nu}=\hat{U}^\dagger\hat{\mathcal{F}}_{\mu\nu}\hat{U}$. There is also the usual U(1) symmetry associated with the ordinary (spin-diagonal) contribution to the vector potential, which describes the magnetic field.

\subsection{Wilson-loop integral}

The $2\times2$ single-electron propagator from $\mathbf{r}_i$ to $\mathbf{r}_f$ along a certain spatial contour $\mathcal{C}$ in time $t$, corresponding to the Hamiltonian (\ref{H}), is given by
\begin{equation}
\hat{K}(\mathbf{r}_f,\mathbf{r}_i;t)=T_\mathcal{C}\,e^{-(2\pi i/\phi_0)\int_\mathcal{C}d\mathbf{r}\cdot\hat{\mathbf{A}}(\mathbf{r})}K(\mathbf{r}_f,\mathbf{r}_i;t)\,.
\label{K}
\end{equation}
Here, $K(\mathbf{r}_f,\mathbf{r}_i;t)$ is the spin-diagonal propagator if we set $\hat{\mathbf{A}}\equiv0$ and $T_\mathcal{C}$ is the path-ordering operator along the contour $\mathcal{C}$. If we are interested in the interference along different trajectories, the basic building block for describing SO coupling effects is provided by the Wilson-loop integral:
\begin{equation}
\hat{W}_\mathcal{C}(\mathbf{r})=T_\mathcal{C}\,e^{-(2\pi i/\phi_0)\oint_\mathcal{C}d\mathbf{r}^\prime\cdot\hat{\mathbf{A}}(\mathbf{r}^\prime)}\,,
\label{W}
\end{equation}
which, for general spatially inhomogeneous systems, is a function of the position $\mathbf{r}$ and the closed contour $\mathcal{C}$, starting and ending at $\mathbf{r}$. Notice that $\hat{W}_\mathcal{C}$ is a purely geometric object dependent on the contour but independent of how fast the particle moves around it. Furthermore, to clarify our nomenclature, this object is called a Wilson-loop integral rather than simply a Wilson loop, because the trace operation has not been performed.

We shall demonstrate in the next section that the theory simplifies either in the presence of strong SO spatial inhomogeneity, making it effectively Abelian to the leading order in the SO strength,\cite{tserkovPRB07sh} or in the case of a simple geometry, such as a single-loop Aharonov-Casher effect.\cite{tserkovPRB07ac} The limit of mesoscopic systems small on the scale of the spin-precession length is also simple from our point of view, since only the leading non-Abelian corrections to the geometric spin transformation need to be retained. Fortunately, furthermore, we will argue that the physics of weak localization also does not require the full non-Abelian treatment in many cases of practical interest. Besides, the weak localization provides an illuminating example, where we can easily relate the Wilson-loop perspective at SO effects with readily measurable quantities.

\section{The Non-Abelian Corrections}
\begin{figure*}[pth]
\centerline{\includegraphics[width=0.8\linewidth]{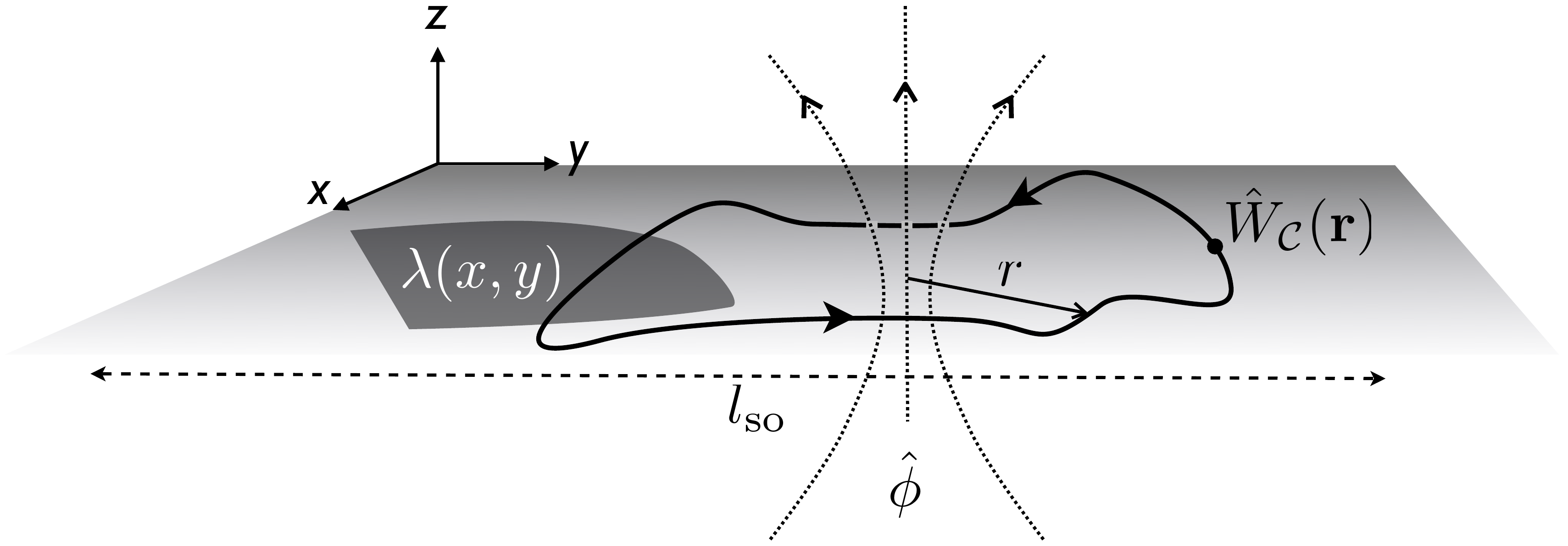}}
\caption{A schematic of the considered two-dimensional model: A 2DEG residing in the $xy$ plane experiences a position-dependent SO coupling parametrized by $\lambda(x,y)$. The dark patch shows a region where the SO strength is enhanced, e.g., by an electric gate or mechanical strain. The SU(2) phase factor accumulated by spin precession in the SO field is represented by the path- and origin-dependent Wilson-loop integral $\hat{W}_\mathcal{C}(\mathbf{r})$, Eq.~(\ref{W}). In the small-loop approximation, Eq.~(\ref{Wa}), the Wilson-loop integral is directly related to the Yang-Mills curvature ``flux" $\hat{\phi}$ through the loop. The curvature (\ref{F}) is dominated by the derivative terms, $\partial_\mu\hat{A}_\nu-\partial_\nu\hat{A}_\mu$, in the regions where the SO strength $\lambda(x,y)$ is rapidly varying (e.g., close to the edges of the dark patch), while its non-Abelian nature is reflected in the commutator contribution, $[\hat{A}_\mu,\hat{A}_\nu]$, in the regions where the SO strength is smooth or constant. The small-loop approximation requires that $r\ll l_{\rm so}$, where $r$ is the characteristic size of the loop. The weak-localization transport corrections near some point $\mathbf{r}$ are governed by the interference between counter-propagating closed trajectories starting and ending at $\mathbf{r}$.}
\label{fig}
\end{figure*}

The Wilson-loop integral (\ref{W}) provides a gauge-covariant description of spin precession and interference in the presence
of an SO interaction. In the case of an Abelian theory, such as U(1) electromagnetism, we can employ the regular Stokes' theorem for evaluating Eq.~(\ref{W}) as follows:
\begin{equation}
\hat{W}_\mathcal{C}(\mathbf{r})=e^{-2\pi i\phi/\phi_0}~~~{\rm (Abelian)}\,,
\label{WA}
\end{equation}
where $\phi=\int_s d\mathbf{S}\cdot(\boldsymbol{\nabla}\times\mathbf{A})$ is the magnetic flux integrated over the oriented surface area $d\mathbf{S}$. Note that the Stoke's theorem, Eq.~(\ref{WA}), provides the manifestly gauge-invariant form for $\hat{W}_\mathcal{C}$, expressed in terms of the magnetic field $\mathbf{B}=\boldsymbol{\nabla}\times\mathbf{A}$.
On the other hand, the non-Abelian Stokes' theorem is required for the SU(2) theory, which relates the curvature given by Eq.~(\ref{F}) to the exponentiated loop integral given by Eq.(\ref{W}), resulting in the following surface integral:
\begin{equation}
\hat{W}_\mathcal{C}(\mathbf{r})=T_{s}\,e^{-(\pi i/\phi _{0})\int_{s}dr^{\prime\mu}\wedge dr^{\prime\nu}\tilde{\mathcal{F}}_{\mu\nu}(\mathbf{r}^\prime)}\,.
\label{ST}
\end{equation}
Here, $T_s$ is the surface-ordering operator \cite{brodaCHA01} and $\tilde{\mathcal{F}}_{\mu\nu}$ is the path-dependent curvature, defined by
\begin{equation}
\tilde{\mathcal{F}}_{\mu\nu}(\mathbf{r}^\prime)=\hat{K}(\mathbf{r},\mathbf{r}^\prime)\hat{\mathcal{F}}_{\mu\nu}(\mathbf{r}^\prime)\hat{K}(\mathbf{r}^\prime,\mathbf{r})\,.
\end{equation}
Clearly the mathematical structure for non-Abelian fields requires a more elaborate approach than the familiar Abelian version, and one has to contend with an object that is more complex than a simple magnetic flux. A more general 2D theory would require evaluating Eq.~(\ref{ST}) over many possible closed trajectories, which is a formidable task without a simple representation for the surface integral on the right-hand side of Eq.~(\ref{ST}). Therefore, as a first step, let us examine some asymptotic limits within this framework. For simplicity, we will take specific case of the Rashba form, corresponding to
\begin{equation}
\hat{\mathbf{A}}(\mathbf{r})=\lambda(\mathbf{r})\,\mathbf{z}\times\hat{\boldsymbol{\sigma}}\,,
\label{AR}
\end{equation}
where $\mathbf{r}$ lies in the $xy$ plane, $\mathbf{z}$ is a normal unit vector and $\hat{\boldsymbol{\sigma}}$ is a vector of the Pauli matrices, which are generators of the SU(2) group.  A schematic of our model is shown in Fig.~\ref{fig}.

\subsection{Homogeneous SO coupling}
\label{hsoc}

As the SO coupling strength $\lambda(\mathbf{r})$ becomes more homogeneous, the derivative components of Eq.~(\ref{F}) vanish and the curvature field is determined entirely by the commutator: $\hat{\mathcal{F}}_{12}\propto i[\hat{A}_1,\hat{A}_2]\propto\hat{\sigma}_z$. Consider the integration in the exponent of Eq.~(\ref{W}).
For simplicity, we can take the closed contour $\mathcal{C}$ to be a unit square loop in the $xy$ plane. For a uniform square, connecting the points $(0,0)\to(1,0)\to(1,1)\to(0,1)\to(0,0)$, $\hat{W}_\mathcal{C}=e^{i\hat{A}_y}e^{i\hat{A}_x}e^{-i\hat{A}_y}e^{-i\hat{A}_x}$ (absorbing the $2\pi/\phi_0$ factor by the redefinition of the connection $\hat{\mathbf{A}}$). Since the Cartesian components of $\hat{\mathbf{A}}$ are noncommutative, the concatenated exponents must be expanded in the Baker-Campbell-Hausdorff series, which to the leading nontrivial order in SO strength becomes: 
\begin{align}
\hat W_\mathcal{C}&=e^{i\hat{A}_y} e^{i\hat{A}_x}e^{-i\hat{A}_y}e^{-i\hat{A}_x}\nonumber\\
&=e^{i\hat{A}_x+i\hat{A}_y+[{\hat{A}_x ,\hat{A}_y}]/2+\cdots}e^{-i\hat{A}_x-i\hat{A}_y+[{\hat{A}_x,\hat{A}_y}]/2+\cdots}\nonumber\\
&\approx e^{[{\hat{A}_x,\hat{A}_y}]}\,. 
\end{align}

In this approximation, $\hat{W}_\mathcal{C}(\mathbf{r})$ can be rewritten in the form of the Abelian Stokes' theorem as (restoring the $2\pi/\phi_0$ factor):
\begin{equation}
\hat{W}_\mathcal{C}\approx e^{-2\pi i\hat{\phi}/\phi _0}\,,
\label{Wa}
\end{equation}
in terms of the ``plain" flux of the field $\hat{\mathcal{F}}$ through the loop in 2D, without any surface ordering or path dependence,
\begin{equation}
\hat{\phi}=\frac{1}{2}\int_{s}dr^{\prime\mu}\wedge dr^{\prime\nu}\hat{\mathcal{F}}_{\mu\nu}(\mathbf{r}^\prime)=\int_{s}dS\,\hat{\mathcal{F}}_{12}(\mathbf{r}^\prime)\,.
\label{Fp}
\end{equation}
For the Rashba model (\ref{AR}), this becomes\cite{tserkovPRB07ac}
\begin{equation}
\hat{\phi}=-(4\pi\lambda^2S/\phi_0)\hat{\sigma}_z=-\pi\phi_0(r/l_{\rm so})^2\hat{\sigma}_z\,,
\label{fR}
\end{equation}
where $l_{\rm so}=\phi_0/2\lambda$ is the spin-precession length. The relative correction to the plain flux in the exponent of Eq.~(\ref{Wa}) scales as $r/l_{\rm so}$ (irrespective of the detailed shape of the loop), so that the approximation requires that $r\ll l_{\rm so}$.

\subsection{Strongly Inhomogeneous SO coupling}
\label{isoc}

For strong spatial variations in $\lambda(\mathbf{r})$, the first term of Eq.~(\ref{F}) dominates, and such a variation of $\lambda(\mathbf{r})$ results in the fictitious magnetic field,\cite{tserkovPRB07sh}
\begin{equation}
\hat{\mathbf{B}}=\boldsymbol{\nabla}\times\hat{\mathbf{A}}=\left(\hat{\boldsymbol{\sigma}}\cdot\boldsymbol{\nabla}\lambda\right)\,\mathbf{z}\,,
\label{B}
\end{equation}
directed along the $z$ axis. With respect to the spin space, the fictitious field $\hat{\mathbf{B}}$ has opposite signs for spins up and down along the gradient $\boldsymbol{\nabla}\lambda$. In Ref.~\onlinecite{tserkovPRB07sh}, the problem of the ``boundary spin Hall effect" was discussed where $\lambda(\mathbf{r})$ changes abruptly from some constant value $\lambda_0$ to zero along the edge of a 2D Hall bar. This problem is of interest as a candidate model for a lateral Hall contact for spin injection by a Rashba system.\cite{adagideliPRL05} If the SO strength $\lambda$ variation is abrupt on the scale of $l_{\rm so}$, the fictitious field (\ref{B}) dominates the boundary physics, resulting in a spin-dependent (ordinary) Hall effect. The noncommutative contribution to the gauge field (\ref{F}) in this case is small.

The lengthscale characterizing the SO inhomogeneity is given by
\begin{equation}
l_{\rm inh}\sim\lambda/|\boldsymbol{\nabla}\lambda|\,,
\end{equation}
which has to be shorter than $l_{\rm so}$ for the gauge structure to reduce to the essentially Abelian theory. In such strongly inhomogeneous limit, local (on the scale of $l_{\rm so}$) SO physics is dominated by the fictitious field (\ref{B}), while the non-Abelian commutator contribution to the curvature (\ref{F}) becomes relatively unimportant. Invoking the Abelian Stokes' theorem, we thus reduce the problem to the approximation (\ref{Wa}), where the flux is now given by the familiar relation:
\begin{equation}
\hat{\phi}=\int_sd\mathbf{S}\cdot\hat{\mathbf{B}}=\int_s dS(\hat{\boldsymbol{\sigma}}\cdot\boldsymbol{\nabla}\lambda)\,.
\label{fI}
\end{equation}

By comparing the limits of the homogeneous vs strongly inhomogeneous SO interactions of the Rashba model, one discovers one conspicuous difference: In the former case, the spin quantization axis set by the curvature ``flux" (\ref{fR}) is along the normal $z$, while in the latter it is along the in-plane direction determined by the gradient $\boldsymbol{\nabla}\lambda$, Eq.~(\ref{fI}).\cite{tserkovPRB07ac} In the very special case of a combination of the linear Dresselhaus and Rashba SO interactions of equal strength, we have an Abelian SU(2) field with zero curvature in homogeneous systems.\cite{bernevigPRL06}. In this trivial limit, the SO coupling can be eliminated by an appropriate gauge transformation. 

\subsection{Small-loop limit}

Both the homogeneous and strongly inhomogeneous limits considered above reduced to the same approximation, Eq.~(\ref{Wa}), in terms of the $2\times2$ flux $\hat{\phi}$ of the curvature tensor component $\hat{\mathcal{F}}_{12}$. The difference between the two limits was only in which was the dominant contribution to the curvature (\ref{F}): the derivative, $\partial_1\hat{A}_2-\partial_2\hat{A}_1$, or the commutator, $[\hat{A}_1,\hat{A}_2]$, piece. In particular, in both of these special cases, we were not concerned with the surface ordering required by the non-Abelian Stokes' theorem, Eq.~(\ref{ST}), as long as spins precess and interfere on lengthscales less than the characteristic $l_{\rm so}$. In the opposite limit, the fully non-Abelian topological structure should become manifest.

Let us notice that the integral (\ref{W}) enters naturally in the semiclassical path-integral construction of the quantum transport corrections.\cite{chakravartyPRP86} In the next section, we will demonstrate that Eq.~(\ref{Wa}) gives an adequate approximation for many cases of interest, in this context. The physical reasoning behind this is as follows: The onset of the non-Abelian effects as the loop size approaches $l_{\rm so}$ also signals the onset of destructive spin interference between time-reversed trajectories and thus suppression of quantum corrections stemming from the larger loops. The singlet Cooperon channel, which is invariant under SO precession, provides an exception to this argument.

\section{transport corrections in a 2DEG}

\subsection{Preliminaries}
We start by recalling the path-integral formalism behind the semiclassical treatment of weak localization and the DC transport in disordered systems. Let us consider a system of noninteracting electrons moving in a random environment. Clearly, any deviations from an ideal crystal will result in scattering events that will contribute to the resistivity. The classical paths of the scattered electrons can be represented by Brownian motion. In a 2DEG, the electrons will move around the material with group velocities close to the Fermi velocity $v_F$. Furthermore, in the limit of a large system size, the particles that experience a random walk will also obey Fick's law for diffusion, $\mathbf{J}=-D\boldsymbol{\nabla}\rho$, where the ordinary diffusion constant in two dimensions is:
\begin{equation}
D = v_F^2\tau/2=l^2/2\tau\,,
\end{equation}
for a given mean free path $l$ and scattering time $\tau$. The conductivity $\sigma$ and $D$ are related by the Einstein relation $\sigma  = e^2 N(0)D$, where $N(0)$ is the density of states at the Fermi level. It should be emphasized that in the semiclassical approach, $D$ can be computed explicitly from the velocity-velocity correlation function, which can be understood as characterizing the time $t$ it takes for an electron to forget its initial velocity direction. Since we are interested in the contributions to the velocity-velocity correlation function that are responsible for quantum corrections to the DC transport, we focus on the set of the Brownian paths the electrons will take and the associated probabilities for propagation and return.

Weak localization is a quantum-interference correction to $\sigma$ that results from phase-coherent backscattering. Consequently, these corrections follow from a modified diffusion model that includes interference effects. Any additional phenomena modifying the weak localization itself must be corrections affecting time-reversal properties of electronic trajectories, which can destroy the phase coherence of back-scattered waves. In particular, since the singlet Cooperon channel is responsible for the weak antilocalization, suppressing the triplet channels may delocalize electrons. The most commonly used formalism in weak localization calculations is usually taken in the language of impurity averaged propagators and maximally crossed Langer-Neal diagrams. However, a semiclassical approach is more intuitive and  physically appealing, as the effect can be understood in a more transparent description as a modified diffusion process yielding the following quantum correction to the classical conductivity $\sigma$:\cite{chakravartyPRP86}
\begin{equation}
\sigma_Q=-\frac{4e^2}{h}D\int_\tau^\infty dt\tilde{R}(t)e^{-t/\tau_\varphi}\,.
\label{gQ}
\end{equation}
Here, $dt\tilde{R}(t)$ is the return probability associated with the interference of all classical paths with their time-reversed counterparts within the time interval $(t,t+dt)$. The elastic scattering time $\tau$ serves as the lower limit of integration, while the upper limit is effectively set by the phase-coherence time $\tau_\varphi\gg\tau$, according to the factor $e^{-t/\tau_\varphi}$.
Furthermore, the explicit averaging over the random potential can be carried out by incorporating random fluctuations into the classical paths. 
The classical probability of return can be formally written as $R(t)=\int d[\mathbf{r}(t^\prime)]P_t[\mathbf{r}(t^\prime)]$, for closed paths in a neighborhood of the trajectory $\mathbf{r}(t^\prime)$. The probability for the realization of a (coarse-grained) Boltzmannian path $\mathbf{r}(t^\prime)$ is given by the Wiener measure:
\begin{equation}
P_t[\mathbf{r}(t^\prime)]\propto e^{-\int_0^tdt^\prime\dot{\mathbf{r}}(t^\prime)^2/4D}\,.
\label{Pt}
\end{equation}
In the presence of a SO interaction (\ref{H}), the interference between counter-propagating closed trajectories acquires an SU(2) phase correction (\ref{W}). For a sufficiently weak SO interaction, the trajectories are assumed to be otherwise unaffected. (We will specify the exact condition below.) The interference between the counter-propagating closed trajectories is affected by the spin precession as follows:\cite{chakravartyPRP86}
\begin{equation}
\hat{R}(t)=\int d[\mathbf{r}(t^\prime)]P_t[\mathbf{r}(t^\prime)]\hat{W}_\mathcal{C}[\mathbf{r}(t^\prime)]\hat{W}_\mathcal{C}^\dagger[\bar{\mathbf{r}}(t^\prime)]\,,
\end{equation}
where $\bar{\mathbf{r}}(t^\prime)$ is the time-reversed trajectory. The return probability $\tilde{R}$ entering Eq.~(\ref{gQ}) has to be appropriately spin-averaged: $\tilde{R}={\rm Tr}[\hat{R}]/2$. Let us notice that $\hat{W}_\mathcal{C}^\dagger[\bar{\mathbf{r}}(t^\prime)]=\hat{W}_\mathcal{C}[\mathbf{r}(t^\prime)]$ and use the approximation (\ref{Wa}) to find for the spin return interference:
\begin{equation}
\hat{W}_\mathcal{C}[\mathbf{r}(t^\prime)]\hat{W}_\mathcal{C}^\dagger[\bar{\mathbf{r}}(t^\prime)]\approx e^{-4\pi i\hat{\phi}/\phi_0}\,.
\label{Wc}
\end{equation}
We will return later to discussing the range of validity of this approximation in the context of weak localization. Tracing this over spin to get the quantum conductivity correction (\ref{gQ}), we see that the calculation now reduces to finding the eigenstates $\pm\phi$ of the traceless Hermitian matrix $\hat{\phi}$, in the case of an SU(2) field. For the more general SU(2)$\times$U(1) field (e.g., Rashba SO plus the ordinary electromagnetic vector potential), $\hat{\phi}$ acquires a finite trace, with the two eigenstates $\phi_\pm$ becoming
\begin{equation}
\phi_\pm=\phi_m\pm\phi\,,\,\,\,\hat{\phi}=\phi_m+\phi\,\hat{\boldsymbol{\sigma}}\cdot\mathbf{n}\,.
\label{phi}
\end{equation}
Here, $\phi_m$ is the ordinary magnetic flux through the loop, $\phi$ is the SO contribution, and $\mathbf{n}$ is a unit vector defining spin-quantization axis. In general, by not relying upon the approximation (\ref{Wa}), the same composition is upheld, although the Hermitian matrix $\hat{\phi}$ that determines spin precession for a given loop is not given simply by the flux (\ref{Fp}).\cite{tserkovPRB07ac} Note that we have neglected Zeeman splitting, focusing on the case when the magnetic field is oriented perpendicular to the plane of the electron motion.\cite{maekawaJPSJ81}

It is well known that weak localization is affected by the presence of magnetic fields and SO interactions. Both effects modify electronic diffusion and subsequently, the phase coherence between time reversed paths. In the case of a magnetic field, phase coherence is destroyed in a process called ``anomalous magnetoconductance." Within our Abelian approximation (\ref{Wa}), the intrinsic SO quantum corrections to the conductivity can be expressed in the same language as magnetoconductance due to a spin-dependent fictitious magnetic field. However, because of the different effect on the singlet and triplet Cooperon channels, a sufficiently strong SO interaction can reverse the weak localization leading to antilocalization, and the conductivity that would decrease with an increasing magnetic field. In this case, the relative spin orientation of the interfering electron trajectories is important. In particular, in certain special cases of interest, it can be possible to localize spin species asymmetrically, with the corresponding spin-quantization axis tunable by a combination of the strength and the gradients of the SO coupling parameter $\lambda(\mathbf{r})$. We call this limit ``spin-selective localization."

\subsection{Spin-selective localization}

One intriguing consequence of Eq.~(\ref{phi}) is the possibility to delocalize electrons spin-selectively by an appropriate combination of an applied magnetic field and an adjustable SO interaction. The relevant spin-quantization axis $\mathbf{n}$ can, for example, be chosen to be along the in-plane gradient of a strongly inhomogeneous SO strength, Eq.~(\ref{fI}). This provides a pragmatic scenario, since we could have large fictitious fields with moderate magnitudes of the SO interaction, while also allowing for the in-plane freedom in choosing the spin-quantization axis. In practice, however, this hinges on the ability to control large gradients of the SO strength with elastic strain or electric gates, for instance.

The quantum conductivity correction in the presence of an effective flux (\ref{phi}) is given in 2D for each spin by \cite{chakravartyPRP86}
\begin{equation}
\sigma_Q^\pm=-\psi\left(\frac{1}{2}+\frac{\phi_0}{8\pi|\phi_\pm|}\frac{\tau_\varphi}{\tau}\right)+\psi\left(\frac{1}{2}+\frac{\phi_0}{8\pi|\phi_\pm|}\right)\,,
\label{psi0}
\end{equation}
in terms of the digamma function $\psi$, in units of $e^2/2\pi h$. When $|\phi_\pm|\ll\phi_0$, this approximates to
\begin{equation}
\sigma_Q^\pm\approx-\ln\frac{\tau_\varphi}{\tau}+\frac{2}{3}\left(\frac{2\pi \phi_\pm}{\phi_0}\right)^2~~~(|\phi_\pm|\ll\phi_0)\,.
\label{psi}
\end{equation}
$\phi_\pm$ entering these equations is given by the flux of the effective field $B_\pm=B_m\pm B$ through the area $l_\varphi^2=D\tau_\varphi$. $B_m$ is the out-of-plane magnetic field, and the fictitious field $\pm B$ is given by the eigenvalues of the curvature
\begin{equation}
\hat{\mathcal{F}}_{12}=\hat{\boldsymbol{\sigma}}\cdot\boldsymbol{\nabla}\lambda-(\pi\phi_0/l_{\rm so}^2)\hat{\sigma}_z\,,
\label{F12}
\end{equation}
assuming for simplicity a constant gradient $\boldsymbol{\nabla}\lambda$, on the scale of $l_\varphi$. Notice that because of the covariance of the Yang-Mills curvature $\hat{\mathcal{F}}$, $B_\pm$ must be gauge invariant. In the opposite limit of $|\phi_\pm|=|B_\pm|l^2_\varphi\gg\phi_0$ (when, e.g., $\tau_\varphi\to\infty$), Eq.~(\ref{psi0}) approximates to
\begin{equation}
\sigma_Q^\pm\approx-\ln\frac{\phi_0}{8\pi|B_\pm|D\tau}~~~(|\phi_\pm|\gg\phi_0)\,.
\label{gQa}
\end{equation}
The above limits are obtained easily by using the asymptotic behavior of the digamma function: $\psi(1/2+z)=\ln z+1/24z^2+\mathcal{O}(1/z^4)$ at $z\to\infty$ and $\psi(1/2+z)=\mathcal{O}(z)$ at $z\to0$.

It is necessary to remark that our semiclassical treatment based on Eqs.~(\ref{Pt})-(\ref{Wc}) requires that $|B_\pm|\ll\phi_0/l^2$ (where $l=v_F\tau$ is the mean free path), so that we always have $(\phi_0/|\phi_\pm|)(\tau_\varphi/\tau)\gg1$. In particular, for the field $B_\pm$ determined by the SO curvature (\ref{F12}), this translates into the requirement $l\ll l_{\rm so}$, when $\lambda$ is uniform. Note that for a fixed and finite SO interaction, Eq.~(\ref{psi}) corresponds to a dip of the total conductivity as a function of the physical magnetic field $B_m$ at zero field, since $(\phi_+^2+\phi_-^2)/2=\phi^2+\phi_m^2$, while Eq.~(\ref{gQa}) has a peak at $B_m=0$, since $|\phi_+\phi_-|=|\phi^2-\phi_m^2|$. [See, however, the discussion below leading to Eq.~(\ref{lQ}).]

It is now appropriate to discuss the legitimacy of the approximation (\ref{Wa}), in the context of the weak localization corrections. As mentioned earlier, this approximation requires that $r\ll l_{\rm so}$. Otherwise, the omitted non-Abelian corrections to the Stokes' theorem due to spin precession become appreciable. Let us turn off the  magnetic field ($\phi_m=0$) and return to the two extreme cases discussed in Secs.~\ref{hsoc} and \ref{isoc}: Homogeneous $\lambda$, such that the curvature field $\hat{\mathcal{F}}_{12}$ is determined by the second term in Eq.~(\ref{F12}), and large constant gradient $\boldsymbol{\nabla}\lambda$, such that the curvature is dominated by the first term. In the former case, the flux through the area $l_{\rm so}^{2}$ is $\phi\sim\lambda^2l_{\rm so}^2/\phi_0\sim\phi_0$, which means the non-Abelian corrections to the preceding weak-localization analysis based on Eq.~(\ref{Wa}) can become appreciable if $l_{\rm so}\ll l_\varphi$. In fact, we in general cannot reduce the problem to uncorrelated propagation of two spin projections, since only the triplet subspace of the Cooperon precesses on the scale of $l_{\rm so}$, while the singlet channel is unaffected by SO coupling.\cite{chakravartyPRP86,skvortsovJETPL98} Eq.~(\ref{gQa}) corresponding to $l_{\rm so}\ll l_\varphi$, in the case of a homogeneous Rashba SO with $\phi_m=0$, thus should only be valid for the triplet channels, for which $l_{\rm so}$ sets the cutoff lengthscale for trajectories contributing to the coherent backscattering. Separating the antilocalization singlet contribution, we thus immediately find
\begin{equation}
\sigma_Q/2\approx\ln(l_\varphi/l)-3\ln(l_{\rm so}/2\pi l)~~~(l_\varphi\gg l_{\rm so})\,,
\label{lQ}
\end{equation}
in agreement with Ref.~\onlinecite{skvortsovJETPL98}. The factor of 3 accounts for the triplet degeneracy. In particular, note the crossover from localization to antilocalization, as $l_{\rm so}$ is made shorter. In the other extreme of a large spin-orbit gradient, we should generally have a similar concern of non-Abelian corrections beyond Eq.~(\ref{Wa}): Although the larger gradient increases the curvature field and shrinks the relevant lengthscale corresponding to $\phi\sim\phi_0$, it is easy to show the spin precession will remain significant. We will not pursue this problem in detail here, but it is worthwhile to remark that it can be overcome, e.g., by tuning the Rashba parameter $\lambda$ in combination with the linear Dresselhaus SO interaction, to reduce the non-Abelian contribution to the curvature field.\cite{bernevigPRL06}

\subsection{Intrinsic vs extrinsic SO coupling}

It is also instructive to compare the limit of weak intrinsic SO coupling, Eq.~(\ref{psi}), to the delocalizing correction in the presence of a weak extrinsic (random) SO interaction due to quenched disorder:\cite{chakravartyPRP86}
\begin{equation}
\sigma_Q/2\approx-\ln(\tau_\varphi/\tau)+2\tau_\varphi/\tau_{\rm so}\,,
\end{equation}
where $\tau_{\rm so}^{-1}$ is the extrinsic SO scattering rate due to impurities. In the case of a homogeneous Rashba SO coupling (\ref{AR}), the D'yakonov-Perel (DP) spin-relaxation rate \cite{zuticRMP04} is $\tau_{\rm DP}^{-1}\propto\lambda^2\tau$, in the relevant regime of $l\ll l_{\rm so}$. The leading SO correction to the localization (\ref{psi}) is thus proportional to $\tau_{\rm DP}^{-2}$, i.e., the square of the DP spin-relaxation rate, since the curvature defining the effective flux $\phi_\pm$ is proportional to $\lambda^2$, for a uniform $\lambda$.

We can easily understand the different localization dependence on the spin-relaxation rates, $\tau^{-1}_{\rm so}$ and $\tau^{-1}_{\rm DP}$, in the two cases, by examining how the SU(2) phase prefactor in the propagator (\ref{K}) depends on the strength of the SO coupling for closed loops. The uncorrelated quenched SO disorder leads to a memoryless (Markovian) spin precession, which is similar for both open and closed trajectories. However, the intrinsic SO coupling (\ref{AR}) combined with ordinary momentum scattering results in a DP spin-precessional random walk, which is qualitatively different for open and closed paths. In the case of open trajectories responsible for DP spin relaxation, this random walk is Markovian. However, the net spin precession is reduced by closing a trajectory, because $\langle\hat{\mathbf{A}}\cdot d\mathbf{r}/dt\rangle_{\rm loop}=0$, hence the higher-order scaling of the delocalizing correction with the spin-relaxation rate. Note that this vanishing of the average field driving spin precession along the closed loops is exact only for a homogeneous $\lambda$. Making $\lambda$ inhomogeneous would enhance the delocalizing corrections, since now $\langle\hat{\mathbf{A}}\cdot d\mathbf{r}/dt\rangle_{\rm loop}\neq0$. For strongly inhomogeneous $\lambda$, the curvature (\ref{F12}) scales linearly with the SO strength $\lambda$, and the delocalizing correction linearly with the DP spin-relaxation rate $\tau_{\rm DP}^{-1}$. 

This is analogous to the effect of SO interaction on weak localization and quantum conductance fluctuations in a chaotic quantum dot:\cite{brouwerPRB02} The relevant SO scattering rate is substantially reduced due to the geometrical constraint on particle trajectories through a chaotic quantum dot, in the case of a uniform SO vector potential (\ref{AR}). By introducing SO nonuniformity, however, the rate of SO scattering can be augmented to a level comparable to that in the bulk.\cite{brouwerPRB02} SO inhomogeneities are in fact inevitable even in high-mobility quantum wells due to fluctuations in the concentration of remote dopant ions.\cite{shermanAPL03} Their important role has been established experimentally \cite{mullerPRL08} with regard to spin relaxation in symmetric quantum wells, where the ordinary D'yakonov-Perel mechanism is inactive. The nonexponential spin relaxation due to random SO field in the presence of magnetic field has also been recently proposed,\cite{glazovPRB05} noting, in particular, qualitative difference between the open and closed trajectory contributions. These interesting effects remain to be explored in the context of quantum corrections to transport.

\section{Discussion}

Let us finally discuss some potential practical consequences of controlling the curvature field (\ref{F12}) by tuning the appropriate SO coupling profile $\lambda(\mathbf{r})$. The local SO strength $\lambda$ determines the curvature component $\propto\lambda^2$ with the spin-quantization axis out of plane, while the gradient $\boldsymbol{\nabla}\lambda$ controls the in-plane curvature component. The combination of the two determines the net spin-quantization axis $\mathbf{n}$, while the eigenvalues $\pm B$ of the curvature matrix $\hat{\mathcal{F}}_{12}$, which are opposite for the two spins, combined with the ordinary magnetic field $B_m$, which is the same for the two spins, can in principle result in any desirable spin-dependent field $B_\pm$. One interesting possibility would be to tune the field $B_\pm$ deep in the localized regime such that it vanishes for one spin species only, say $B_-=0$. In this case, only the spin $\uparrow$ would be delocalized by a large enough field $B_+$, along a desirable spin-quantization axis. This could pave the way for a gate- or strain-tunable spin filter in two dimensions, without the use of ferromagnetic materials, which may become useful for developing semiconductor-based spintronic applications.

\begin{acknowledgments}
We thank Arne Brataas, Sudip Chakravarty, and Xun Jia for numerous helpful discussions. This work supported in part by the Alfred P. Sloan Foundation (YT). 
\end{acknowledgments}

\end{document}